\documentstyle[a4,12pt,axodraw]{article}
\def\be{\begin{equation}}
\def\ee{\end{equation}}
\def\bq{\begin{eqnarray}}
\def\eq{\end{eqnarray}}
\begin{document}

\thispagestyle{empty}
\setcounter{page}{0}

\begin{flushright}
CERN-TH.6880/93\\
MPI-Ph/93-32\\
LMU-07/93\\
May 1993
\end{flushright}
\vspace*{\fill}
\begin{center}
{\Large\bf QCD Calculation of the $B \rightarrow \pi,K$ Form Factors }$^1$\\
\vspace{2em}
\large
{\bf V.M. Belyaev$^{a}$, A. Khodjamirian$^{b,c*}$, R. R\"uckl$^{b,d,e}$}\\
\vspace{2em}
{$^a$ \small Institute of Theoretical and Experimental Physics, 117259 Moscow,
Russia }\\
{$^b$ \small Sektion Physik der Universit\"at M\"unchen, D-8000 M\"unchen 2,
FRG}\\
{$^c$ \small Yerevan Physics Institute, 375036 Yerevan, Armenia } \\
{$^d$ \small Max-Planck-Institut f\"ur Physik, Werner-Heisenberg-Institut,
D-8000 M\"unchen 40, FRG}\\
{$^e$ \small CERN, CH-1211 Geneva 23, Switzerland}\\
\end{center}
\vspace*{\fill}

\begin{abstract}

We calculate the form factors for the heavy-to-light transitions
$B\rightarrow \pi,K $ by means of QCD sum rules using $\pi$ and $K$
light-cone wave functions. Higher twist contributions as well as gluonic
corrections are taken into account.
The sensitivity to the shape of the leading-twist wave functions and
effects of SU(3)-breaking are discussed. The results are compared with quark
model predictions and with
the results from  QCD sum rules for three-point correlators.

\end{abstract}

\vspace*{\fill}

\begin{flushleft}
\noindent$^1$ Work supported by the German Federal Ministry for Research and
Technology under contract No. 05 6MU93P\\
\noindent$^*$ {\it Alexander von Humboldt Fellow}\\
\end{flushleft}

\newpage

\section{Introduction}

The development of reliable quantitative methods to calculate hadronic matrix
elements of quark and gluon operators in QCD is a task of great theoretical
and phenomenological importance. In particular, the understanding of weak
decays of charmed and beautiful hadrons and their use in determining
fundamental parameters of the standard model and testing the theory depend
crucially on the progress made in solving this difficult problem.

In this paper we consider the heavy-to-light transition form factors
appearing in the matrix elements
$< \pi \mid \bar{q}\gamma_\mu b \mid \bar{B}>$, $q=u,d$  and
$< K \mid \bar{s}\gamma_\mu b \mid \bar{B}>$ which play an important role in
$B$-meson decays. The former of these matrix elements determines CKM
suppressed semileptonic B-decays, while the latter one enters in factorizable
amplitudes of nonleptonic $B$-decays. One of the most interesting examples is
the mode $B\rightarrow J/\psi K$. Theoretically, these form factors represent
relatively simple hadronic matrix elements and are therefore very useful to
test and compare different approaches.

The early  calculations of heavy-to-light $B$-meson form factors were performed
at zero momentum transfer in the framework of a quark model \cite{BSW} as well
as using standard QCD sum rules \cite{DP,ovch}. The dependence on the
four-momentum transfer $p$ was assumed to be given by a simple pole factor
$(1-p^2/m_*^2)^{-1}$, with $m_*$ being the mass of the lowest lying resonance
in a given channel. The first complete calculation of the
$B \rightarrow \pi$ form factor for nonzero momentum transfer was carried out
quite recently \cite{BBD2} using
QCD sum rules for three-point correlators. The results actually support the
pole dominance model.

The aim of the present paper is to call attention to an alternative way of
calculating heavy-to-light form factors which is also based on
short-distance operator product expansion and QCD sum rule methods.
However, in the present approach the basic correlator is taken between
the vacuum and a light meson state, and all information about
large distances is incorporated into a set of so-called light-cone wave
functions for that particular light meson \cite{chernyakpr}. These  wave
functions represent distributions of the light-cone momentum of the
constituents, and can be classified by their twist defined as the
difference between the canonical dimension and the Lorentz spin of
the corresponding operator. The light-cone wave functions are universal
quantities similarly as the vacuum condensates, and are by now quite well
studied. Their asymptotic form is fixed by perturbative QCD, while the
nonasymptotic effects at lower momentum scales can be estimated from QCD sum
rules for two-point correlators of appropriate currents
\cite{chernyakpr,gorsky}.

In refs. \cite{BBK,braun} the light-cone method was demonstrated to be very
suitable for exclusive processes. Moreover, it is technically much simpler
than the standard QCD sum rule approach where all participating hadrons are
replaced by corresponding currents with euclidean momenta.
Previous applications of light-cone sum rules to weak form factors
of heavy mesons focussed on the  $B \rightarrow \pi$ form factor at zero
momentum transfer \cite{cz} and on the $D \rightarrow K $ form factor
\cite{BBD}.

Here, we present and discuss a comprehensive calculation of the
$B \rightarrow \pi,K$ form factors which takes into account all twist 2, 3
and major twist 4 contributions. In addition to the pure quark wave functions,
we also evaluate the influence of quark-gluon wave functions. The sensitivity
to the precise shape of the leading-twist wave function and the impact of
$SU(3)$-breaking effects are investigated in some details.
Finally, we compare our results with quark model \cite{BSW} and conventional
QCD sum rule calculations \cite{BBD2}.

\section {Formal derivation of the  sum rules }

For definiteness, we show here and in the next section the derivation of the
sum rule for the $B^-\rightarrow K^-$ form factor. The results can be
easily generalized to the corresponding expressions for $B \rightarrow \pi$
form factors. We start with the following matrix element of the time-ordered
product of two currents between the vacuum state and the K-meson at momentum
$q$:
\be
F_\mu (p,q)=
i \int d^4xe^{ipx}<K(q)\mid T\{\bar{s}(x)\gamma_\mu b(x),
\bar{b}(0)i\gamma_5 u(0)\}\mid 0> \, .
\label{1a}
\ee

The hadronic representation of (1) is obtained by inserting a complete set
of states including the $B$-meson ground state, higher resonances and
nonresonant states with B-meson quantum numbers:
\be
F_\mu (p,q)=
\frac{<K\mid \bar{s}\gamma_\mu b\mid B>
<B\mid \bar{b}i\gamma_5u\mid 0>}{m_B^2-(p+q)^2}
+
\label{2a}
\ee
$$
+\sum_h \frac{ <K\mid \bar{s}\gamma_\mu b\mid h>
<h\mid \bar{b}i\gamma_5u\mid 0>}{m_h^2-(p+q)^2}
$$
$$
=F(p^2,(p+q)^2)q_\mu +\tilde{F}(p^2,(p+q)^2)p_\mu \, .
$$
Here, $p$ denotes the four-momentum transfer. Otherwise, the notation
should be self-explain\-ing. From now on, we shall concentrate on the invariant
amplitude $F$ which is physically more interesting than the amplitude
$\tilde{F}$. For $F$ one can write a general dispersion relation in the
momentum squared $(p+q)^2$ of the $B$-meson:
\be
F(p^2,(p+q)^2)= \int_{m_B^2}^\infty
\frac{\rho (p^2,s)ds}{s-(p+q)^2}
\label{3a}
\ee
where possible subtractions are neglected and the spectral density
is given by
\be
\rho (p^2,s)=\delta (s-m_B^2)2f^+_K(p^2)\frac{m_B^2f_B}{m_b}+ \rho^{h}(p^2,s)
\,.
\label{4a}
\ee
The first term on the r.h.s. of (4) represents the $B$-meson contribution
and follows from (\ref{2a}) by inserting the matrix elements
\be
<K\mid \bar{s}\gamma_\mu b\mid B>=2f^+_K(p^2) q_\mu +(f^+_K(p^2)+
f^-_K(p^2)) p_\mu
\label{5a}
\ee
and
\be
<B\mid \bar{b}i\gamma_5u\mid 0>=\frac{m_B^2f_B}{m_b} .
\label{fB}
\ee
In the above, $ f^+_K(p^2) $ is the transition form factor for
$B^- \rightarrow K^-$ which we want to calculate,  $f_B$ is the $B$-meson
decay constant, $m_b$ is the $b$-quark mass, and $\rho^h(p^2,s)$
denotes the spectral density of higher resonances and of the continuum of
states. In accordance with the standard procedure in the QCD sum
rule applications, we  invoke quark-hadron duality and replace
$\rho^h$ by
\be
\rho^h(p^2,s)=\frac{1}{\pi}Im F_{QCD}(p^2,s)\Theta (s-s_0)
\label{6a}
\ee
where $ImF_{QCD}(p^2,s)$ is obtained from the imaginary part of the correlation
function (\ref{1a}) calculated in QCD and $s_0$ is the threshold
parameter.
For consistency, we shall take both parameters $ f_B $ and $s_0$
from a QCD sum rule analysis of the correlator of two $ \bar{b}\gamma_5 u $
currents, as explained in Section 3.

In order to suppress the contribution from the excited states and from the
continuum exponentially and to get rid of possible subtraction terms, we apply
the Borel operator $ \hat{B}$ with respect to the variable $(p+q)^2$ to
the dispersion integral in (\ref{3a}) :
\be
F(p^2,M^2) = \hat{B} F(p^2,(p+q)^2) =\int_{m_B^2}^\infty \rho (p^2,s)
e^{ -s/M^2}ds \, .
\label{7a}
\ee
Using then  (\ref{4a}) and (\ref{6a}) , one obtains
\be
F(p^2,M^2)=2\frac{f_Bm_B^2}{m_b}f^+_K(p^2)e^{-m_B^2/M^2}+
\frac{1}{\pi}\int_{s_0}^\infty ImF_{QCD}(p^2,s)e^{-s/M^2}ds \, .
\label{Borel}
\ee
The next step involves the calculation of the correlation function (\ref{1a})
or the invariant amplitude
$F(p^2,(p+q)^2)$ in QCD. This calculation is of
course essentially the same as the one proposed above for (\ref{6a}).
After Borel transformation the result can be written in the form
\be
F(p^2,M^2)=\frac{1}{\pi}\int_{m _b ^2}^\infty ImF_{QCD}(p^2,s)e^{-s/M^2}ds \, .
\label{QCDF}
\ee
Finally , equating (\ref{QCDF}) with (\ref{Borel}) yields the desired sum rule
for the form factor $f^+_K$:
\be
f^+_K(p^2)=\frac{m_b}{2\pi m_B^2f_B}\int_{m_b^2}^{s_0}
ImF_{QCD}(p^2,s)e^{-\frac{s-m_B^2}{M^2} }ds  \, .
\label{8a}
\ee

\section{ QCD calculation with light-cone wave functions}

The possibility to calculate the correlator (\ref{1a}) in the region of large
space-like momenta \\ $(p+q)^2 <0 $, keeping the momentum $q$ at the physical
point $q^2= m_K^2 \simeq 0$, is based on the expansion of the $T$-product of
the currents in (\ref{1a}) near the light-cone $x^2 = 0 $. The leading
contribution to the operator product expansion arises from the contraction
of the b-quark operators in (1) to the free $b$-quark propagator
$ <0 \mid b \bar{b}\mid 0> $. The light quark operators are left uncontracted .
Diagramatically, this
contribution is depicted in Fig. 1a.
The formal expression is easily obtained from  (\ref{1a}):
\be
F_\mu(p,q)=\int_0^\infty
\frac{d\alpha}{16\pi^2\alpha^2}\int d^4xe^{ipx-m_b^2\alpha+
\frac{x^2}{4\alpha}}
\label{9a}
\ee
$$
\times \left(m_b<K(q)\mid\bar{s}(x)\gamma_\mu\gamma_5 u(0)\mid 0>+
\frac{ix^\rho}{2\alpha}<K(q)\mid\bar{s}(x)\gamma_\mu\gamma_\rho \gamma_5
u(0)\mid 0>
\right)
$$
where we made use of the following representation of the free propagator
$\hat{S}_b^0$ :
\be
<0\mid T\{b(x)\bar{b}(0)\}\mid 0>=i\hat{S}_b^0(x)=\int \frac{d^4p}{(2\pi)^4}
e^{-ipx}i\frac{\hat{p}+m_b}{p^2-m_b^2}
\label{10a}
\ee
$$
=-\int_0^\infty \frac{d\alpha}{16\pi^2\alpha^2}(m_b+i
\frac{\hat{x}}{2\alpha})e^{-m_b^2\alpha+
\frac{x^2}{4\alpha}} \,  .
$$
According to the analysis presented in \cite{BBD} the expansion
(\ref{9a}) should be valid in the range $0 \leq p^2 \leq m_b^2-O(1GeV^2)$,
i.e. sufficiently far below the physical states in the $b\bar{s}$-channel.
The lowest lying charmonium levels evidently reside in this
region, so that our results will be applicable to $ B \rightarrow J/\psi K $.

Let us consider the first term in (\ref{9a}). The matrix element of the
nonlocal operator is given by \cite{chernyakpr,braun}:
\be
<K(q)\mid\bar{s}(x)\gamma_\mu\gamma_5u(0)\mid 0>=
-iq_\mu f_K \int_0^1due^{iuqx}\left(
\varphi_K (u)+\frac{5\delta^2_K}{36}x^2\varphi_{4K}(u)\right)
\label{11a}
\ee
where  $\varphi_K(u)$ is the $K$-meson light-cone wave function
of the leading twist 2 and $ \varphi_{4K}(u)$ represents one of the
next-to-leading twist 4 wave functions.
There are also other twist 4 terms in this matrix element specified in
\cite{braun} which we do not show here explicitly since their numerical
contribution to the final result is of the order of $1\%$ as we have
checked.
All wave functions are normalized to unity and  $\delta^2_K$ is a
dimensionful parameter given later.

The matrix element appearing in the second term of eq. (\ref{9a}) can be
split in two matrix elements using the identity $ \gamma_\mu\gamma_\rho=
-i\sigma_{\mu \rho} +g_{\mu\rho} $. These matrix elements are determined
by wave functions of twist 3:

\be
<K(q)\mid \bar{s}(x)i\gamma_5u(0)\mid 0>=
\frac{f_K m_K^2}{m_u+m_s}\int_0^1due^{iuqx}\varphi_{pK}(u)
\label{12a}
\ee

\be
<K(q)\mid\bar{s}(x)\sigma_{\mu\nu}\gamma_5u(0)\mid 0>=
i(q_\mu x_\nu -q_\nu x_\mu )\frac{f_K m_K^2}{6(m_u+m_s)}
\int_0^1due^{iuqx}\varphi_{\sigma K}(u) \, .
\label{12b}
\ee
 Substituting  the matrix elements (\ref{11a}) to (\ref{12b}) into (\ref{9a}),
and integrating over $x$ and the auxiliary parameter $\alpha$ we find the
following
expression for the coefficient of $q_\mu$ in (\ref{2a}) that is for the
invariant amplitude $F$:
\be
F_{QCD}(p^2,(p+q)^2) = -f_Km_b\int_0^1du\left(
\frac{\varphi_K (u)}{(p+qu)^2-m_b^2}
- \frac{10 \delta^2_Km_b^2 \varphi_{4K}(u)}
{9((p+qu)^2-m_b^2)^3}\right)
\label{15a}
\ee
$$
-\frac{f_K m_K^2}{m_u+m_s} \int_0^1du\left[\frac{\varphi_{pK}(u)u}{(p+qu)^2
-m_b^2}+\frac{\varphi_{\sigma K} (u)}{6((p+qu)^2-m_b^2)}
\left(2-\frac{p^2+m_b^2}{(p+qu)^2-m_b^2}\right)\right] \, .
$$

Borel transformation of (\ref{15a}) according to (\ref{7a}) then yields
the QCD result anticipated in (\ref{QCDF}). Finally, the $B \rightarrow K$
transition form factor $f^+_K(p^2)$ is obtained from (\ref{8a})
and (\ref{15a}):
\be
f^+_K( p^2)= \frac{f_K m_b^2}{2f_Bm_B^2}\int_\Delta^1\frac{du}{u}
exp[\frac{m_B^2}{M^2}-\frac{m_b^2-p^2(1-u)}{uM^2}]
\label{20a}
\ee
$$
[ \varphi_K(u) + \frac{\mu_K}{m_b}u\varphi_{pK}(u) + \frac{\mu_K}{6m_b}\varphi_
{\sigma K}(u)(2 + \frac{m_b^2+p^2}{uM^2})
- \frac{5\delta^2_Km_b^2}{9M^4} \frac{\varphi_{4K}(u)}{u^2} ] \, .
$$
where $\mu_K = m_K^2/(m_s+m_u)$ and  $\Delta = (m_b^2-p^2)/(s_0-p^2) $ .

In addition to the quark-antiquark wave functions considered above
there are in principle also contributions from multi-particle wave functions.
The most important corrections of this type are expected
to arise from quark-gluon operators in the operator-product expansion of
(\ref{1a}). A typical diagram where the gluon is emitted from the heavy quark
is shown in Fig.1b. The leading contribution arises from the twist 3 operator:
\newpage
\be
<K(q)\mid \bar{s}(x)gG_{\mu\nu}(z)\sigma_{\rho\lambda}\gamma_5u(0)\mid 0>
=if_{3K}[q_\mu (q_\rho g_{\lambda\nu}-q_\lambda g_{\rho\nu})
\label{16a}
\ee
$$
-q_\nu (q_\rho g_{\lambda\mu}-q_\lambda g_{\rho\mu})]
\int{\cal D}\alpha_i\varphi_{3K}(\alpha_i)e^{iq(x\alpha_1+z\alpha_3)}
$$
where $ G_{\mu\nu}(z)=(\lambda^c/2)G_{\mu\nu}^c(z)$, $\lambda^c$
and $ G_{\mu\nu}^c$ being the usual colour matrices and the gluon field
tensor, and ${\cal D}\alpha_i=d\alpha_1d\alpha_2d\alpha_3\delta
(\alpha_1+\alpha_2+\alpha_3-1)$. The three-particle wave function \\
$\varphi_{3K}( \alpha_i )$ $= \varphi_{3K}( \alpha_1, \alpha_2, \alpha_3)$
and the corresponding coupling constant $ f_{3K} $ are introduced and discussed
in \cite{chernyakpr,gorsky}.

The calculation of this gluonic correction is somewhat more involved than the
previous one. Instead of the free propagator (\ref{10a}) we now need the
b-quark propagator including the interaction with gluons in first order :
\be
<0\mid T\{b(x)\bar{b}(0)\}\mid 0>_A=i\hat{S}_b^0(x)- ig\int dz \hat{S}_b^0(x-z)
\gamma^\mu \frac{\lambda^c}{2}A^c_\mu(z)\hat{S}_b^0(z)
\label{18a}
\ee
where $\hat{S}_b^0(x)$ is the free $b$-quark propagator defined in (\ref{10a}).
In the fixed point gauge, $ x^\mu A^c_\mu = 0 $, the
gluon field $A^c_\mu$ can be represented directly in terms of the field
strength $G_{\mu\nu}^c$:
\be
A_\mu^c(z)= z_\nu\int ^1_0 uduG_{\nu\mu}^c(uz)\, .
\label{44a}
\ee
We substitute (\ref{18a}) into (\ref{1a}), use (\ref{16a}) and (\ref{44a}),
and integrate over $x$ and $z$. This yields the following expression for the
quark-gluon contribution to the invariant amplitude $F$ :
\be
F^G_{QCD}(p^2,(p+q)^2)=4f_{3K}\int_0^1u du\int{\cal D}\alpha_i
\frac{(pq)\varphi_{3K}(\alpha_i)}
{((p+(\alpha_1+u\alpha_3)q)^2-m_b^2)^2} \, .
\label{19a}
\ee
Finally, after Borel transformation of (\ref{19a}) one arrives at
the corresponding correction to $f^+_K$:
\be
f^{+G}_K(p^2) = -\frac{ f_{3K}m_b}{f_Bm_B^2 }\int_0^1u du\int {\cal D}\alpha_i
\Theta( \alpha_1+u\alpha_3-\Delta)
\label{55a}
\ee
$$
exp[\frac{m_B^2}{M^2}-\frac{m_b^2-p^2(1-\alpha_1-u\alpha_3)}{(\alpha_1+
u\alpha_3)M^2}][1-\frac{ m^2_b -p^2 }{(\alpha_1+u\alpha_3)M^2}]
\frac{\varphi_{3K}(\alpha_i)}{(\alpha_1+ u\alpha_3 )^2} \, .
$$

Before concluding the QCD calculation of the $B \rightarrow K $ form factor
$f^+_K(p^2)$ we want to make a few remarks on the remaining gluon corrections.
One can distinguish three types of corrections. There are corrections where a
gluon is emitted by one of the quark lines in Fig. 1a and absorbed in the
meson wave function. One of these contributions is shown in Fig. 1b and
calculated above. The others, when the gluon is emitted by a light quark line
effectively give rise to three-particle corrections to the wave
functions of twist 3 as it was shown in \cite{braun}. In the next section,
we will study the numerical influence of these corrections on $f^+_K$.
Furthermore, there are corrections due to the interaction of the heavy virtual
quark in the correlator (\ref{1a}) with the vacuum gluon condensate. They are
expected to be of order of $< G_{\mu\nu}G_{\mu\nu}>/m_b^4 $ and, therefore
negligible for the $b$-quark. Corresponding effects for the light quarks are
considered to be included  in the light-cone wave functions.
Finally, there are additional perturbative $ O(\alpha_s)$ corrections from
gluon exchange between the light and heavy quarks. These corrections
are beyond the scope of the present paper. However, we will approximately
take into account the important $ O(\alpha_s)$ correction to the pseudoscalar
vertex $ \bar{b}\gamma_5 u$ shown in Fig. 1c following the procedure advocated
in \cite{BBD2}.

In order to obtain the corresponding expressions for the $B \rightarrow \pi $
form factor $f_\pi^+$ from (\ref{20a}) and (\ref{55a}), one only needs to
replace  $f_K $ by  $f_{\pi}$,  $\mu_K$ by $\mu_\pi$,  $\varphi_K $ by
$\varphi_\pi$,  etc.
It is important to note that twist 3 contributions to  $f_K^+$ and  $f_\pi^+$
are suppressed by a factor $\mu_K/m_b $  and $\mu_\pi/m_b $, respectively.
Therefore, one expects the form factors to be dominated by the contributions
from the twist 2 wave functions. This dominance may in fact be exploited to
improve our knowledge of these universal wave functions by comparing
the predictions which will be presented in section 5 with experimental data.

\section{Choice of parameters and wave functions}

Having the necessary formulae at hand we next explain our choice of the
relevant parameters and wave functions.
Let us first consider the parameters which characterize the $B$-meson channel
that is $ m_B= 5.28 GeV $, $f_B$, $m_b$ and $s_0$. We note that what was
actually calculated in (\ref{20a}) is a sum rule for the product of two
amplitudes, $f^+_Kf_B$. On the other hand, the $B$-meson decay constant $f_B$
is determined independently by a well-known QCD sum rule for the two-point
correlator of $\bar{b}\gamma_5u$-currents. As it was stressed in \cite{BBD2},
the consistent and convenient way to deal with the above parameters is to take
their values from this two-point sum rule, but neglect $ O(\alpha_s)$ radiative
corrections in $f_B$ since these corrections tend to be cancelled by the
corresponding radiative corrections to the sum rule (\ref{20a}) depicted in
Fig. 1c .
In other words, instead of substituting the physical value of $f_B$ in
(\ref{20a}) we may take $\tilde{f}_B$ defined as the square root of the
two-point sum rule for $f_B^2$ without radiative gluon corrections, and
at the same time drop the correction from Fig. 1c diagram.
{}From a numerical analysis of the sum rule for $f_B^2$ we find several
self-consistent sets of parameters: $\tilde{f}_B=135 MeV $, $ m_b=4.7 GeV $,
$ s_0=35 GeV^2$ (I), $\tilde{f}_B= 160 MeV$, $ m_b= 4.6 GeV $,
$ s_0=37 GeV^2 $  (II), and $\tilde{f}_B= 115 MeV$, $ m_b= 4.8 GeV$ and
$s_0= 33 GeV^2$ (III). As our nominal choice we take set (I) which is
in accordance with the values used in the conventional QCD sum rule approach
of \cite{BBD2}. However, we have checked that sets (II) and (III) lead to
almost the same results.

A second set of parameters is connected with  the light mesons $\pi$ or $K$.
These are  $f_\pi=133 MeV$, $f_K= 160 MeV$, $ \mu_\pi = m_\pi^2/(m_u+m_d)$
and $ \mu_K = m_K^2/(m_s+m_d)$. With $m_u+m_d \simeq 11 MeV$
\cite{gasserleutwyler} one has $ \mu_\pi \simeq 1.6 GeV$
( at a normalization scale of order $1GeV$). To estimate $\mu_K$ we use the
familiar PCAC relation for pseudo-Goldstone bosons to get
\be
\frac{\mu_K}{\mu_{\pi}} = \frac{(<\bar{u}u> + <\bar{s}s>)f_\pi^2}{(<\bar{u}u>
+ <\bar{d}d>)f_K^2} \simeq 0.62 \, .
\label{Kpcac}
\ee
Here, we assumed  $<\bar{u}u>:<\bar{d}d>:<\bar{s}s> \simeq 1:1:0.8 $ as
suggested by QCD sum rules for strange hadrons (see e.g.
\cite{narison}). With $ \mu_\pi \simeq 1.6 GeV$ one thus obtains
$ \mu_K \simeq 1.0 GeV$. On the other hand, simple $SU(3)$ and $SU(6)$
symmetry arguments suggest $\mu_K \simeq \mu_\pi $, i.e. a considerably larger
value. Interestingly, the latter choice corresponds to
$m_s + m_u \simeq 150 MeV$,
whereas the use of (\ref{Kpcac}) implies a heavier s-quark.
Fortunately, the uncertainty in $\mu_K$ only affects the
twist 3 contribution to the $B\rightarrow K$ form factor, and therefore
does not preclude reasonably accurate predictions.

Finally, for the parameters $ \delta^2_\pi$ and $ \varphi_{3\pi} $ which
appear in the coefficients of the higher twist wave functions
$ \varphi_{4\pi}$ and $ \varphi_{3\pi} $, respectively, we take the values
given in the literature \cite{chernyakpr,gorsky}:
$ \delta^2_\pi \simeq 0.2 GeV^2 $ and $f_{3\pi}\simeq 0.0035GeV^2 $.
Since these contributions are small, our results are quite insensitive to
possible differences between $\pi$ and $K$ in this respect and we simply put
$ \delta^2_K \simeq \delta^2_\pi$ and $ \varphi_{3K} \simeq \varphi_{3\pi} $.

Turning now to the light cone wave functions themselves, we recall
the asymptotic expressions which are completely determined by
perturbative QCD and well-known \cite{chernyakpr}. In the SU(3) limit,
one has ( $H=K,\pi$ )
\be
\varphi_H= 6u(1-u)
\label{13a}
\ee
\be
\varphi_{pH}=1,\hspace{0.3cm} \varphi_{\sigma H}= 6u(1-u)
\label{tw3as}
\ee
\be
\varphi_{4H}= 30u^2(1-u)^2
\label{tw4as}
\ee
\be
\varphi_{3H} =360\alpha_1\alpha_2\alpha_3^2 \, .
\label{17a}
\ee
Clearly, the asymptotic wave functions should be  renormalized to the
characteristic scale of the process under consideration. This brings
nonasymptotic effects into play which change the shape of the above wave
functions, but preserve their normalization to unity. Moreover, SU(3)-breaking
effects give rise to asymmetries in the  K-meson wave functions under
$ u \leftrightarrow 1-u $ reflecting the asymmetry in the quark masses
$m_s$ and $m_{u,d}$. Since in the Breit frame the light-cone wave functions
represent distributions of the fraction of the  $K$-meson momentum carried by
the constituent quarks, one would expect the s-quark to carry more momentum
on average than the light quarks.

Being of nonperturbative origin, these effects are difficult to evaluate.
Fortunately, since the form factor $f^+(p^2)$ only depends on
integrals over the light-cone wave functions, it is not neccessary to know
their actual shape very precisely. Moreover, the Borel mass parameter
$M^2 \geq O(M_B^2-m_b^2)$ provides a reasonably high renormalization
scale. Therefore, one can expect the asymptotic wave functions to yield a quite
reliable estimate. Nevertheless, in order to be on the safe side, we have
investigated nonasymptotic effects for the leading-twist wave functions
$ \varphi_\pi$ and $\varphi_K$. For that purpose we have used the model
suggested in \cite{chernyakpr} which is based on an expansion over orthogonal
Gegenbauer polynomials with coefficients determined by means of QCD sum
rules for the 2-point correlators of $\pi$ and $ K$ currents.
The explicit expressions are given below :

\be
\varphi_K= 6u(1-u)\{1 + A^+_K[(2u-1)^2 -\frac{1}{5}]+5b(2u-1)[1+A^-_K[(2u-1)^2
-\frac{3}{7}]\}
\label{14a}
\ee

\be
\varphi_\pi = 6u(1-u)[1+ A^+_\pi[(2u-1)^2 -\frac{1}{5}]] \, .
\label{pion}
\ee
Taking for the parameters in (\ref{14a}) and (\ref{pion}) the values fitted at
the normalization scale  $ \mu \simeq 500 MeV $ \cite{chernyakpr,cz}
and renormalizing these values to the scale $\mu \simeq M_B^2-m_b^2$ one finds
$A^+_K=1.8$, $A^-_K=1.2$, $A^+_\pi=3.0$ and  $b \simeq 0.1 $. Note that the
parameter $b$  incorporates the difference between the average $s$- and
$u$-quark momenta. We shall study numerically the change in $f^+$ due to
changes in the shape of $\varphi_\pi$ and $\varphi_K$. For the higher-twist
wave functions $\varphi_p$, $\varphi_{\sigma}$,  $\varphi_4 $ the asymptotic
expressions should provide sufficiently reliable estimates of the subleading
contributions.

Concluding the discussion of parameters and wave functions it is important to
note that at the level of accuracy adopted here, there are at least three
sources of SU(3)-breaking effects which may cause differences between the
$B\rightarrow \pi$ and $B\rightarrow K$ form factors: the difference between
$f_\pi$ and $f_K$, the difference between $ \mu_\pi$ and $ \mu_K$, and the
difference between $\varphi_\pi$ and $\varphi_K$. Further refinement in this
respect  is possible, but this is far beyond the goal of our present
study.

\section{Numerical results}

Before giving numerical predictions on the form factors $f^+(p^2)$ we must
first determine the range of values for the Borel parameter $ M^2$ for which
the sum rules (\ref{20a}) can be expected to yield reliable results. The lower
limit of this range is determined by the requirement on the terms proportional
to $M^{-2n}$, $n>1$ to remain subdominant. In (\ref{20a}) this concerns in
particular the twist-4 contribution from $\varphi_4$, which increases
rapidly at small $M^2$ analogously to higher order power corrections in
conventional QCD sum rules. The upper limit  of the allowed interval in $M^2$
is determined by demanding the higher resonance and continuum contribution not
to grow too large. We have checked  numerically for $ 0 \leq p^2 \leq 20 GeV^2$
and $ 10 \leq M^2 \leq 15 GeV^2 $ that
(a) the contribution from $ \varphi_4 $ is less than $10 \% $ ,
(b) the higher states contribute less than $30 \% $ , and
(c) the resulting values of $ f^+(p^2) $ are practically independent of the
Borel parameter at $p^2 \leq 10 GeV^2$ and vary only slightly with $M^2$ at
$p^2 \geq 10 GeV^2$. This is illustrated in Fig. 2 for $f^+_K(p^2)$.
The breakdown of the stability as $p^2$ approaches the region $m_b^2-O(1GeV^2)$
is expected. From the fiducial window in Fig. 2a we can read off the QCD
prediction for the $B\rightarrow K$ form factor at zero momentum transfer:
\be
f^+_K(0)= 0.32 \, .
\label{fk}
\ee
The variation with $M^2$ is about 0.005. The $p^2$-dependence of this
form factor is plotted in Fig. 3.

While the twist 4 contribution from $ \varphi_{4K} $ do not exceed $ 10\% $
as required and the gluonic correction from $\varphi_{3K}$ is found to be
smaller than $2\%$, the twist 3 wave functions contribute at the level of  $20$
to $40\%$ and are therefore important. They are shown separately in Fig. 2.

The stability features and the hierarchy in the contributions from the wave
functions of different twist are essentially the same when going from the
$K$-meson to the $\pi$-meson. For the  $B \rightarrow \pi$ form factor at
zero momentum transfer we find
\be
f^+_{\pi}(0)= 0.29 \, .
\label{fpi}
\ee
Here, the result varies with $M^2$ within the fiducial range by 0.01.
The $p^2$ dependence of this form factor is shown in Fig. 4.

It is of course important to investigate further sources of
theoretical uncertainties. Therefore, we have carefully examined the
sensitivity of $f^+_{K}(p^2)$ and $f^+_{\pi}(p^2)$ to reasonable changes
of the parameters and wave functions discussed in the previous section.
As far as the $B$-channel parameters $f_B$, $ m_b $ and $s_0$ are concerned,
we stress again that they are interrelated through sum rules for two-point
correlators. Hence, they should not be changed independently. Numerically, for
the three consistent sets of values given in section 4 , the resulting form
factor vary by less than $5\%$ despite the sizable variation of the individual
parameters. Obviously, $f_B$ enters only in the total normalization of the
form factors and drops out in the ratio
$f^+_{\pi}(p^2) /f^+_{K}(p^2)$, whereas $m_b$
also sets the main scale for the $p^2$-dependence.

In contrast, the sensitivity to the parameters $\mu_\pi$ and $\mu_K$ entering
the coefficients of the twist 3 contributions is quite considerable. This
causes no problem for $f^+_{\pi}(p^2)$ since $\mu_\pi$ is known with good
accuracy. However,  this is not the case for $\mu_K$ as pointed out in
section 4 where we have obtained values varying from 1.0 to 1.6 GeV.
The corresponding uncertainty in $f^+_{K}(p^2)$ amounts to $20\%$ at large
$p^2$ as illustrated in Fig. 5a.

Finally, we have studied the influence of the shape  of the
leading-twist wave functions $ \varphi_\pi $ and $ \varphi_K $ which  give
the dominant contributions. In Fig. 5b, $f^+_{K}(p^2)$
is compared for two different choices for $\varphi_K$ :
(a) the asymmetric form  (\ref{14a}) used in all numerical illustration
presented so far, and (b) the asymptotic form given in (\ref{13a}).
The analogous comparison for $f^+_{\pi}(p^2)$ is shown in Fig. 5c.
As can be seen, the nonasymptotic corrections damp the increase of the
form factor with $p^2$. However, this effect is much more pronounced in
$B \rightarrow \pi$ than in $B \rightarrow K $. On the other hand,
the absolute value of $f^+_{\pi}(p^2)$ remains almost unchanged,
whereas the value of $f^+_K(p^2)$ grows by roughly 0.1 at $p^2\leq 10 GeV^2$.

Some corrections are also expected from nonasymptotic effects in the twist 3
wave functions, including SU(3)-breaking effects. As far as we know, the
latter have not been considered so far. In the SU(3)-limit, nonasymptotic
corrections to the wave functions $\varphi_p $ and $\varphi_\sigma$
are investigated in \cite{braun}. It is shown that they are determined
by the wave function $\varphi_{3K}$ given in (\ref{16a}) above.
As already pointed out, these corrections can be associated with gluons
emitted from the light quark lines in the Fig. 1a and absorbed in the
meson wave function. Using explicit formulae from \cite{braun}, we illustrate
the effect on $f^+_K$ and $f_\pi^+$ in Fig. 5b and c. Clearly, in order
to eliminate the uncertainties illustrated in Fig. 5 a better understanding of
the higher twist wave functions is required.

\section {Conclusions}

Summarizing our investigations, in Figs. 3 and 4 we compare our predictions on
$f^+_K(p^2)$ and $f^+_{\pi}(p^2)$ with the results of other calculations
\cite{BSW,BBD2,cz}. Within the uncertainties there is satisfactory agreement.
To be more definite, at zero momentum transfer we find $f^+_K(0) = 0.26 \div
0.37$ and $f^+_\pi(0)= 0.24 \div 0.29$ where the ranges indicate our estimate
of theoretical uncertainties on the basis of  Fig. 5. We would like to
emphasize in particular the coincidence with the result
$f^+_\pi(0) = 0.24 \pm 0.025 $ obtained in \cite{BBD2} from an alternative
QCD sum rule approach in which the large-distance effects are parametrized in
terms of vacuum condensates rather than by wave function on the light-cone.

Also the $p^2$-dependence of the form factors is rather similar in the
different approaches. Note, however, that in the quark-model \cite{BSW}
the form factors are assumed to  have a simple pole-behaviour:
\be
f^+(p^2) = \frac{f^+(0)}{1-p^2/m_*^2}
\label{21a}
\ee
with $m_*= 5.3 GeV$ in the case of $f^+_\pi$ and $m_*= 5.43 GeV$ for
$ f^+_K$ as expected in the spirit of vector dominance. The authors of
\cite{BBD2} have also presented their calculated result for $f^+_\pi$
in the form (\ref{21a}) and obtained $m_* = 5.2 \pm 0.05 GeV$.
In comparison to that we find a slightly steeper $p^2$-dependence
corresponding roughly to $m_* \simeq 5.0 GeV$.

Furthermore, it is obvious from Figs. 3 to 5 that the method put forward
here is not yet precise enough to deal with $SU(3)$-breaking effects. In other
words, we are not in a position to clearly distinguish
$f^+_K(p^2)$ from $f^+_{\pi}(p^2)$, apart from the tendency
to get a slightly smaller value for $f^+_{\pi}(0)$ than
for $f^+_K(0)$. Improvements in this direction are possible, but require
some further study.

On the other hand, we are able to present a quite accurate prediction for the
value of the form factor $f^+_K$ at the $J/\psi$ mass, to wit
\be
f^+_K( m_{J/\psi}^2)= 0.50 \div 0.60  \, .
\label{psi}
\ee
This result is important, since it is obtained from a well-defined and, as we
have shown, rather stable calculation. It puts investigations of the
$ B \rightarrow J/\psi K $ decay mode, in particular, in the context of
possible searches for $CP$-violation, on a more reliable basis.

Last but not least, the use of light-cone wave functions simplifies the
calculation of weak matrix elements considerably in comparison to the
conventional QCD sum rule approach. This may become crucial in more
complicated problems such as calculations of exclusive nonleptonic B-decays
beyond the factorization approximation.

\section{Acknowledgements}
We are grateful to V.M. Braun for very useful discussions and suggestions. The
help of F. Cuypers in computer calculations is greatly acknowledged.
V.B. thanks the Max-Planck Institute for hospitality and financial support
during the initial phase of this work. A.K. gratefully acknowledges financial
support from the Alexander von Humboldt Foundation.

\pagebreak

\newpage
\begin{center}
{\bf Figure Captions}
\end{center}
\vspace{1cm}

{\bf Figure 1}: QCD diagrams contributing to the correlation function
(1). Solid lines represent quarks, dashed lines gluons, wavy
lines are external currents, and the blobs denote $K$-meson wave functions
on the light-cone.

\vspace{2cm}

{\bf Figure 2 }: Form factor  $f^+_{K}(p^2)$ as a function of the
Borel mass squared $M^2$ at various values of the momentum transfer $p^2$.
The solid curves depicts the total sum rule results,
while the dashed curves show the twist 3 contribution alone. The arrows
indicate the fiducial interval in $ M^2$.

\vspace{2cm}

{\bf Figure 3 }: Form factor $f^+_{K}(p^2)$ of the  $B \rightarrow K $
transitions at $M^2=10GeV^2$ (upper solid curve) and
$M^2=15GeV^2$ (lower solid curve) for the nominal choice of parameters and
wave functions specified in section 4. The dash-dotted curve shows the
quark model prediction given in [1] .

\vspace{2cm}

{\bf Figure 4 }: Form factor  $f^+_{\pi}(p^2)$ of $B \rightarrow \pi $
transitions at $M^2=10GeV^2$ (upper solid curve) and  $M^2=15GeV^2$
(lower solid curve) for the nominal choice of parameters and
wave functions as in Fig. 3. The quark model prediction from [1] (dash-dotted
curve)
and the QCD sum rule result from [4] (dashed curve) are shown for
comparison. The arrow indicates the result of a QCD calculation [9] similar
to ours at zero momentum transfer.

\vspace{2cm}

{\bf Figure 5 }:  Sensitivity of the form factors to parameters and wave
functions:(a) $f^+_K$ for $\mu_K = 1.0 GeV$ (solid curve) and
$1.6GeV$ ( dashed curve); (b) $f^+_{K}$ for the nonasymptotic
(solid curve) and asymptotic (dashed curve) twist 2 wave function
$\varphi_K $, and for nonasymptotic corrections included in the twist 3 wave
functions $ \varphi_p$ and $\varphi_\sigma $ (dash-dotted curve);
(c) the same as (b) for $f^+_{\pi}$.

\end{document}